\documentclass[twocolumn,floats,showpacs,superscriptaddress,pre]{revtex4}
\usepackage{amssymb}
\usepackage{graphicx}
\usepackage{dcolumn}
\usepackage{amsmath}
\usepackage{bm}
\usepackage{epsfig}

\def\defi{{\buildrel \;def\; \over =}}

\newcommand{\be}{\begin{equation}}
\newcommand{\ee}{\end{equation}}

\newcommand{\media}[1]{\langle #1 \rangle}

\begin{document}

\title{First- and second-order phase transitions in Ising models on small world networks,
simulations and comparison with an effective field theory}

\author{A. L. Ferreira}
\affiliation{Departamento de F{\'\i}sica and I3N, Universidade de
Aveiro, 3810-193 Aveiro, Portugal} 

\author{J. F. F. Mendes}
\affiliation{Departamento de F{\'\i}sica and I3N, Universidade de
Aveiro, 3810-193 Aveiro, Portugal} 

\author{M. Ostilli }
\affiliation{Departamento de F{\'\i}sica and I3N, Universidade de
Aveiro, 3810-193 Aveiro, Portugal} 
\affiliation{ Statistical
Mechanics and Complexity Center (SMC), INFM-CNR SMC, Italy}

\begin{abstract}
We perform simulations of random Ising
models defined over small-world networks and we check the validity
and the level of approximation of a recently proposed effective
field theory. Simulations confirm a rich scenario with the
presence of multicritical points with first- or second-order phase
transitions. In particular, for second-order phase transitions,
independently of  the dimension $d_0$ of the underlying lattice,
the exact predictions of the theory in the paramagnetic regions,
such as the location of critical surfaces and correlation
functions, are verified.
Quite interestingly,
we verify that the Edward-Anderson model with $d_0=2$ is not
thermodynamically stable under graph-noise.
\end{abstract}

\pacs{05.50.+q, 64.60.aq, 64.70.-p, 64.70.P-}
\maketitle

\email{ostilli@roma1.infn.it}

\section{Introduction} \label{intro}
Disordered systems represent one of the most important fields of
statistical mechanics. Disorder is at the base of many interesting
phenomena whose understanding is often far from being trivial or
immediate. Its use varies from applications in condensed matter
physics to computer science and, more recently, to the broad range
of natural, artificial, and social networks. One of the most
important analytical tool to study disordered systems is
represented by suitable mean-field theories. Originally developed
to understand spin glass models, the replica method (RM) and the
cavity method (CM) are nowadays largely used in many other fields
of statistical physics and computer science and represent the most
powerful analytical methods to investigate the intricate nature of
the spin glass and other disordered phases
\cite{Parisi,MezardP,Diluted}.

Among disordered models whose dimensionality - in a broad sense -
can be considered infinite \cite{MOIII}, one can single out a
``hierarchical'' family of models of increasing difficulty such
as: fully connected models (corresponding to infinite connectivity
in the thermodynamic limit) like the Sherrington-Kirkpatrick model
\cite{SK}, finite connectivity models like the Viana-Bray model
\cite{VB}, and models defined over small-world networks
\cite{Watts,Barrat}. This latter class of models has been
introduced in recent years and represents an important development
in modeling more realistic situations in which the spins, besides
interacting through a random finite connectivity, distributed
according to some given distribution with average degree $c$,
interact also through short-range connections. In other words,
small-world models constitute an interplay between purely random
and regular finite-dimensional models. It is known that, despite
the underlying finite dimensionality $d_0$ present in these kind
of graphs, in the thermodynamic limit, models defined on them
manifest a mean field behavior. This fact, far from being trivial,
to be rigorously proved, may lead to hope that the RM or the CM
could be used to solve small-world models. Indeed such methods
have been already successfully exploited in
\cite{Niko,Niko2,Bolle} for $d_0=1$ small-world models defined
upon adding a Poisson distributed random connectivity $c$ to the
underlying regular one-dimensional chain. However, if we take a
look at the mathematical structure of these methods we recognize
the following. For what concerns the RM, we need to know
analytically the two leading eigenvectors of the transfer matrix
of the Ising model without the shortcuts but immersed in a random
external field; whereas for what concerns the CM, it is essential
that the underlying graph $\mathcal{L}_0$ had a tree-like
structure, \textit{i.e.}, no loops, at least in the thermodynamic
limit. As a consequence, both  methods seem hardly applicable to
small-world models if $d_0>1$. The effective field theory we have
recently developed in \cite{SW}, based on mapping a generic random
model onto a non random one (the ``pure model'')
\cite{MOI,MOII,MOIII} is instead applicable to these models. In
fact, though this theory is able to give exact answers only in the
paramagnetic (P) regions, there is no limitation in the underlying
dimension $d_0$. Whereas the theory can be fully treated
analytically for $d_0\leq 1$, for $d_0>1$, we can still apply it
semi-analytically~\footnote{An exception is the spherical model
which can be treated analytically for any $d_0$; see \cite{SW}.}.
All we need to apply the theory is to solve - analytically or
numerically - the pure model in $d_0$ dimension in the absence of
the random shortcuts and in the presence of a uniform external
field. The values of a certain observable $\mathrm{O}_0$ so
obtained \textit{una tantum} will be then used to get the
corresponding value $\mathrm{O}$ for the model in the presence of
the shortcuts and for any choice of the disorder parameters
(couplings and connectivity). This feature, together with the fact
that the effective field equations of the theory have a very
simple structure and a more immediate physical interpretation
compared to the equations of the RM method (in which the
introduction of several coupled auxiliary fields is necessary),
makes this effective field theory particularly interesting to all
those applications in which $d_0>1$ or else the number of
parameters of the model is high. Of course one has to pay such an
advantage with the impossibility to get exact results out of the P
region. However, as we shall show in this paper, also in the other
regions of the phase diagram the theory succeeds in giving
effective approximations allowing us to obtain important insights
on the frozen states, even if we do not have a direct access to
them, as instead the RM or the CM could do, if they were applicable
also to models with loops.

In this paper we consider random Ising models defined over
small-world networks having an underlying regular lattice
$\mathcal{L}_0$ of dimension $d_0=1,2,3$.
Given the initial lattice $\mathcal{L}_0$ with $N$ sites,
we build the small-world network by adding $cN/2$ links
uniformly spread over $\mathcal{L}_0$. This implies that at each
site, besides the $2d_0$ neighbors, there are additional
long-range neighbors whose number is distributed according
to a Poisson distribution with average $c$.
Interactions act via a coupling $J_0$ for the $2d_0$ short-range neighbors
and via a further coupling $J$ for the other long-range neighbors.
This way of building a
small-world network is different from the re-wiring method of
Watts and Strogatz\cite{Watts} (in which the number of shortcuts per site is also Poissonian distributed)
and it is more convenient for analytical calculations.
However, just by using the effective field theory at the base of this paper,
it is possible to deal with the similar re-wiring small-world models as well, and to show rigorously
that the critical behavior of the two kind of models is identical \cite{SF}.

By using Monte Carlo (MC) simulations we check the predictions of
the effective field theory for the critical surfaces, the
susceptibility, the average magnetization and the two point
connected correlation function as a function of the Euclidean
distance $r$ defined on $\mathcal{L}_0$.

The ferromagnetic Ising model (both $J$ and $J_0$ positive) on
small-world networks has been extensively studied \cite{Barrat,
gitterman, pekalski, lopes,herrero2002,Hastings2}. However, of remarkable
interest, for both its theoretical and practical implications, is
the case with negative short-range antiferromagnetic coupling,
$J_0<0$. The anti-ferromagnetic Ising model on small world
networks was studied in \cite{herrero2008} for the case where
there is only one antiferromagnetic coupling constant ($J_0=J$),
but, apart from the fully connected case \cite{Skantzos}, no
attention has been paid to the fact that, in the more general
situation, may exist multicritical points with first- and
second-order phase transition. In fact, when $J_0<0$ and $J>0$,
the effective field theory predicts two critical temperatures with
first or second-order phase transitions separating two P regions.
In this case, simulation results show large fluctuations and a
rather slow approach to the thermodynamic limit, confirming the
slow dynamics of the frustrated system and the importance to have
analytical or semi-analytical frustrated-free tools to investigate
these models. Quite interestingly, the model with given couplings
$J_0$ and $J$, and slightly different values of the connectivity
$c$, can show drastically different phase diagrams either showing
two ferromagnetic (F) phase transitions or a single spin-glass
(SG) phase transition. In the case of first-order phase
transitions the magnetization discontinuity is not exactly
predicted by the theory but simulation results show clearly the
signature of a discontinuity for both the F and SG order
parameters in correspondence of the theoretical P-SG critical
temperature. Furthermore, we see good agreement of the
susceptibility predictions in the P phase. We also consider a
two-dimensional modified Edwards-Anderson model\cite{edwards} with
added long-range shortcuts. This is a special case where besides
the connectivity disorder there is also disorder on the value of
the short-range coupling. As predicted by the theory, simulations
confirm that any infinitesimal addition of shortcuts leads the
system to have a finite temperature P-SG phase transition which
coincides with the theoretical one.
%

The paper is organized as follows. In Secs. II and III we recall
the definition of the small-world models and the method, which
mainly consists in finding the solution of the self-consistent
equation (\ref{THEOa}) and minimizing the effective free energy
(\ref{THEOll}). In Sec. IV we report our MC simulations for
several interesting cases selected for comparison with the
theoretical predictions. Finally, in Sec. V some conclusions
are drawn.

\section{Small world models}
\label{models}
We consider random Ising models
constructed by super-imposing random graphs with finite average connectivity
onto some given lattice $\mathcal{L}_0$ whose set of bonds $(i,j)$
and dimension will be indicated by $\Gamma_0$ and $d_0$, respectively.
Given an Ising model - shortly \textit{the pure model} -
of $N$ spins coupled over $\mathcal{L}_0$
through a coupling $J_0$ and
with Hamiltonian
\begin{eqnarray}
H_0\defi -J_{0}\sum_{(i,j)\in \Gamma_0}\sigma_{i}\sigma_{j}-h\sum_i \sigma_i
\label{H0},
\end{eqnarray}
and given an ensemble $\mathcal{C}$
of unconstrained random graphs $\bm{c}$, $\bm{c}\in\mathcal{C}$,
whose bonds are determined by the adjacency matrix elements $c_{i,j}=0,1$,
we define the corresponding small-world model
- shortly \textit{the random model} -
as described by the following Hamiltonian
\begin{eqnarray}
\label{H}
H_{\bm{c};\bm{J}}\defi H_0-\sum_{i<j} c_{ij}{J}_{ij}\sigma_{i}\sigma_{j},
\end{eqnarray}
the free energy $F$ and the averages $\overline{\media{\mathop{O}}^l}$
being defined in the usual (quenched) way as
\begin{eqnarray}
\label{logZ}
-\beta F\defi \sum_{\bm{c}\in\mathcal{C}} P(\bm{c})\int d\mathcal{P}
\left(\{{J}_{i,j}\}\right)
\ln\left(Z_{\bm{c};\bm{J}}\right),
\end{eqnarray}
and (in the following a bar notation $\bar{\cdot}$ indicates the two independent averages over the
graph and couplings realizations)
\begin{eqnarray}
\label{O}
\overline{\media{\mathop{O}}^l}\defi
\sum_{\bm{c}\in\mathcal{C}} P(\bm{c}) \int d\mathcal{P}\left(\{{J}_{i,j}\}\right)
\media{\mathop{O}}^l, \quad l=1,2
\end{eqnarray}
where $Z_{\bm{c};\bm{J}}$ is the partition function of the
quenched system

\begin{eqnarray}
\label{Z}
Z_{\bm{c};\bm{J}}= \sum_{\{\sigma_{i}\}}
e^{-\beta H_{\bm{c};\bm{J}}\left(\{\sigma_i\}\}\right)},
\end{eqnarray}
$\media{\mathop{O}}_{\bm{c};\bm{J}}$ the Boltzmann-average of the
quenched system (note that $\media{\mathop{O}}_{\bm{c};\bm{J}}$
depends on the given realization of the ${J}$'s and of $\bm{c}$:
$\media{\mathop{O}}=\media{\mathop{O}}_{\bm{c};\bm{J}}$; for
shortness we will often omit to write these dependencies)

\begin{eqnarray}
\label{OO}
\media{\mathop{O}}\defi \frac{\sum_{\{\sigma_i\}}\mathop{O}_{\bm{c};\bm{J}}e^{-\beta
H_{\bm{c};\bm{J}}\left(\{\sigma_i\}\right)}}{Z_{\bm{c};\bm{J}}},
\end{eqnarray}
and $d\mathcal{P}\left(\{{J}_{i,j}\}\right)$ and $P(\bm{c})$ are
two product measures given in terms of two normalized measures
$d\mu(J_{i,j})\geq 0$ and $p(c_{i,j})\geq 0$, respectively:
\begin{eqnarray}
\label{dP}
d\mathcal{P}\left(\{{J}_{i,j}\}\right)\defi \prod_{(i,j),i<j}
d\mu\left( {J}_{i,j} \right),
\quad \int d\mu\left( {J}_{i,j} \right)=1,
\end{eqnarray}
\begin{eqnarray}
\label{Pg}
P(\bm{c})\defi \prod_{(i,j),i<j} p(c_{i,j}),
\quad \sum_{c_{i,j}=0,1} p(c_{i,j})=1.
\end{eqnarray}
The variables
$c_{i,j}\in\{0,1\}$ specify whether a ``long-range'' bond between the sites
$i$ and $j$ is present ($c_{i,j}=1$) or absent ($c_{i,j}=0$), whereas
the $J_{i,j}$'s are the random variables of the given bond $(i,j)$.
For the $c_{i,j}$'s, we shall consider the following
distribution
\begin{eqnarray}
\label{PP}
 p(c_{ij})=
\frac{c}{N}\delta_{c_{ij},1}+\left(1-\frac{c}{N}\right)\delta_{c_{ij},0},
\end{eqnarray}
where $c>0$.
This choice leads in the thermodynamic limit $N\to\infty$ to a
number of long range connections per site distributed according
to a Poisson law with mean connectivity $c$.

In this paper for the $J_{i,j}$'s we will assume  the distribution
\begin{eqnarray}
\label{dPF}
\frac{d\mu\left( {J}_{i,j} \right)}{d{J}_{i,j}}=\delta\left( {J}_{i,j}-J\right),
\end{eqnarray}

For the short-range nearest-neighbor coupling, $J_0$ we will
consider the distribution,
\begin{equation}
\label{J0cte}
 \frac{d\mu\left(J_0 \right) }{dJ_0}=\delta (J_0-a),
\end{equation}
except in the last case studied, the modified Edwards-Anderson
model, where we consider,
\begin{equation}
\label{J0bimodal} \frac{d\mu_0(J_0)}{dJ_0}=\frac{1}{2}
\delta(J_0-a)+\frac{1}{2} \delta(J_0+a).
\end{equation}

\section{An effective field theory}
Depending on the temperature T, and on the parameters of the
probability distributions, $d\mu$ and $p(c_{i,j})$, the random
model may stably stay either in the paramagnetic (P), in the
ferromagnetic (F), or in the spin glass (SG) phase. In our
approach for the F and SG phases there are two natural order
parameters that will be indicated by $m^{(\mathrm{F})}$ and
$m^{(\mathrm{SG})}$. Similarly, for any correlation function,
quadratic or not, there are two natural quantities indicated by
$C^{(\mathrm{F})}$ and $C^{(\mathrm{SG})}$, and that in turn will
be calculated in terms of $m^{(\mathrm{F})}$ and
$m^{(\mathrm{SG})}$, respectively. To avoid confusion, it should
be kept in mind that in our approach, for any observable
$\mathcal{O}$
there are - in principle - always
two solutions that we label as F and SG,
but, for any temperature,
only one of the two solutions is stable and useful
in the thermodynamic limit.

In the following, we will use the label $\mathop{}_0$ to specify
that we are referring to the pure model with Hamiltonian
(\ref{H0}).
Let $m_0(\beta J_0,\beta h)$ be the stable magnetization of the
pure model with coupling $J_0$ and in the presence of a
uniform external field $h$ at inverse temperature $\beta$. Then,
the order parameters
$m^{(\Sigma)}$, $\Sigma$=F,SG,
satisfy the following self-consistent decoupled equations
\begin{eqnarray}
\label{THEOa}
m^{(\Sigma)}=m_0(\beta J_0^{(\Sigma)},
\beta J^{(\Sigma)}m^{(\Sigma)}+\beta h),
\end{eqnarray}
where the effective couplings $J^{(\mathrm{F})}$, $J^{(\mathrm{SG})}$,
$J_0^{(\mathrm{F})}$ and $J_0^{(\mathrm{SG})}$ are given by
\begin{eqnarray}
\label{THEOb}
\beta J^{(\mathrm{F})}= c\int d\mu(J_{i,j})\tanh(\beta J_{i,j}),
\end{eqnarray}
\begin{eqnarray}
\label{THEOc} \beta J^{(\mathrm{SG})}= c\int
d\mu(J_{i,j})\tanh^2(\beta J_{i,j}).
\end{eqnarray}
For a constant short-range coupling distributed as in
(\ref{J0cte})
\begin{eqnarray}
\begin{array} {l l l}
\label{THEOd} J_0^{(\mathrm{F})}&=& a \\
& & \\
 \beta J_0^{(\mathrm{SG})}&=& \tanh^{-1}(\tanh^2(\beta a)).
 \end{array}
\end{eqnarray}
and for the bimodal distribution (\ref{J0bimodal}),
\begin{eqnarray}
\label{THEOdb}
\begin{array} {l l l}
J_0^{(\mathrm{F})}&=& 0 \\
& & \\
\beta J_0^{SG} &=&  \tanh^{-1}(\tanh^2(\beta a))
 \end{array}
\end{eqnarray}

For the correlation functions we have ${{C}}^{(\Sigma)}$, $\Sigma$=F,SG, where
\begin{eqnarray}
\label{THEOh}
{{C}}^{(\Sigma)}=
{{C}}_0(\beta J_0^{(\Sigma)},\beta J^{(\Sigma)} m^{(\Sigma)}+\beta h)+\mathop{O}\left(\frac{1}{N}\right),
\end{eqnarray}
where ${{C}}_0(\beta J_0,\beta h)$ is the correlation function of
the pure model.
For the corrective $\mathop{O}(1/N)$ term in Eq. (\ref{THEOh}) we remind the
reader to Eq. (33) of \cite{SW}.
Let us indicate by $C^{(\mathrm{1})}$ and $C^{(\mathrm{2})}$
the averages and the quadratic averages over the disorder
of the correlation function of degree, say $k$.
Then, $C^{(\mathrm{1})}$ and $C^{(\mathrm{2})}$, are related to
$C^{(\mathrm{F})}$ and $C^{(\mathrm{SG})}$, as follows
\begin{eqnarray}
\label{THEOa0}
C^{(\mathrm{1})}&=&C^{(\mathrm{F})}, \quad \mathrm{in~F}, \\
\label{THEOa01}
C^{(\mathrm{1})}&=& 0, \quad k ~ \mathrm{odd}, \quad \mathrm{in~SG}, \\
\label{THEOa02}
C^{(\mathrm{1})}&=&C^{(\mathrm{SG})}, \quad k ~ \mathrm{even}, \quad \mathrm{in~SG},
\end{eqnarray}
and
\begin{eqnarray}
\label{THEOa03}
C^{(\mathrm{2})}&=&\left(C^{(\mathrm{F})}\right)^2, \quad \mathrm{in~F}, \\
\label{THEOa04}
C^{(\mathrm{2})}&=&\left(C^{(\mathrm{SG})}\right)^2, \quad \mathrm{in~SG}.
\end{eqnarray}

In particular, for the susceptibility $\tilde{\chi}^{(\Sigma)}$ of
the random model we have:
\begin{eqnarray}
\label{THEOchie}
\tilde{\chi}^{(\Sigma)}=
\frac{\tilde{\chi}_0\left(\beta J_0^{(\Sigma)},\beta J^{(\Sigma)}m^{(\Sigma)}+\beta h\right)}
{1-\beta J^{(\Sigma)}\tilde{\chi}_0
\left(\beta J_0^{(\Sigma)},\beta J^{(\Sigma)}m^{(\Sigma)} +\beta h\right)},
\end{eqnarray}
where $\tilde{\chi}_0$ stands for the susceptibility $\chi_0$ of
the pure model divided by $\beta$ (we will adopt throughout
this dimensionless definition of the susceptibility) and similarly
for the random model.
For the case $\Sigma=$F without disorder ($d\mu(J')=\delta(J'-J)dJ'$ and $d\mu_0(J_0)=\delta(J_0-a)dJ_0$),
Eq. (\ref{THEOchie}) was already derived in \cite{Hastings2} by series expansion
techniques at zero field ($h=0$) in the P region (where $m=0$).

Among all the possible stable
solutions of Eqs. (\ref{THEOa}), in the thermodynamic
limit, for both $\Sigma$=F and $\Sigma$=SG,
the true solution $\bar{m}^{(\Sigma)}$, or leading solution,
is the one that minimizes $L^{(\Sigma)}$
where
\begin{eqnarray}
\label{THEOll}
L^{(\Sigma)}(m)\defi
\frac{\beta J^{(\Sigma)}\left(m\right)^2}{2}+
\beta f_0\left(\beta J_0^{(\Sigma)},\beta J^{(\Sigma)}m+\beta h\right),
\end{eqnarray}
$f_0(\beta J_0,\beta h)$ being the free energy density in the
thermodynamic limit of the pure model with coupling $J_0$
and in the presence of an external field $h$, at inverse
temperature $\beta$. A necessary condition for a solution
$m^{(\Sigma)}$ to be the leading solution is the stability
condition:
\begin{eqnarray}
\label{THEOgen}
{\tilde{\chi}_0\left(\beta^{(\Sigma)}J_0^{(\Sigma)},\beta J^{(\Sigma)}m^{(\Sigma)}+\beta h\right)}
\beta^{(\Sigma)}J^{(\Sigma)}<1.
\end{eqnarray}

For the localization and the reciprocal stability between the F and SG phases
we remind the reader to Sec. IIID of \cite{SW}. We recall however that,
at least for lattices $\mathcal{L}_0$
having only loops of even length, the stable P region is always that
corresponding to a P-F phase diagram, so that in the P region
the correlation functions must be calculated only
through Eqs. (\ref{THEOa0}) and (\ref{THEOa03}).

The inverse critical temperature $\beta_c^{(\Sigma)}$
is solution of the following exact equation
\begin{eqnarray}
\label{THEOg}
{\tilde{\chi}_0\left(\beta_c^{(\Sigma)}J_0^{(\Sigma)},0\right)}
\beta_c^{(\Sigma)}J^{(\Sigma)}=1, \quad \beta_c^{(\Sigma)}<\beta_{c0}^{(\Sigma)},
\end{eqnarray}
where $\beta_{c0}^{(\Sigma)}$ is the inverse critical temperature
of the pure model with coupling $J_0^{(\Sigma)}$. When
$J_0>0$, the constrain in Eq. (\ref{THEOg}) ensures the uniqueness of
the solution. However, if $J_0<0$, Eq. (\ref{THEOg}) in general
admits either 0 or at least 2 solutions (in principle also 4, 6,
etc...).

We end this section by stressing that this method is exact in all
the P region and, at least for second-order phase transitions,
provides the exact critical surface, behavior and percolation
threshold, and that, in the absence of frustration, the order
parameters $m^{(\Sigma)}$ become exact also in the limit $c\to
0^+$, in the case of second-order phase transitions, and in the
limit $c\to\infty$ (see Sec. IIIC of \cite{SW}). Note also that
the order parameters $m^{(\Sigma)}$, and then the correlation
functions, are by construction always exact in the zero
temperature limit.

\section{Simulations and comparison with the theory for given couplings}
The Monte-Carlo simulations presented in this work were made using
a local spin-flip dynamics with a Metropolis acceptance
probability \cite{metropolis}.

Throughout this work we estimate the susceptibility, in the
P phase  by $\tilde{\chi}=N \overline{\left < m^2
\right> }$, and in the ferromagnetic phase by $\tilde{\chi}=N
\left (\overline{\left < m^2 \right
>} - \overline{\left < |m| \right>}^2 \right)$, where $m=\frac{1}{N}\sum_{i=1}^N \sigma_i$  is the magnetization of the
system and $N=L^{d_0}$  is  the total number of spins in the
lattice of side $L$.  The Binder cumulant\cite{Binder}, defined by
\begin{eqnarray}
\label{BinderCumulant} U_L=1-\frac{\overline{\left < m^4\right
>}}{3 \overline{\left
< m^2\right >}^2} ,
\end{eqnarray}
was used to locate the critical points. The cumulants $U_L$ and
$U_{L'}$, for two systems of different sides $L$ and $L'$, plotted
as a function of temperature, cross at the critical point at a
value, $U^*$ that characterizes the universality class of the
model.

To study spin-glass phases we calculate the overlap
order-parameter, $q=\sum_{i=1}^N \sigma_i^{(1)} \sigma_i^{(2)}$
obtained from two replicas of the system with spins
$\sigma_i^{(1)}$ and $\sigma_i^{(2)}$. The observed distribution
of the values of $q$ is measured for a given realization of the
disorder which corresponds to taking a thermal average.
Subsequently, by considering different samples,  an average over
disorder is done:

\begin{equation}
P(q)=\overline{\frac{1}{M} \sum_{j=1}^M \delta_{q_j,q}}
\end{equation}
where $q_j$ is the value of the overlap parameter at time step j,
$\delta_{i,j}$ is the Kronecker delta, $M$ is the number of
simulation Monte Carlo steps (MCS) after thermal equilibration is
reached, and the bar denotes averaging over disorder. The Binder
cumulant for the overlap order parameter can be defined by
\begin{equation}
\label{BinderCumulantOverlap} U_{q,L}=1-\frac{ \overline {\left
<q^4\right >}}{3 \left (\overline{\left < q^2\right >} \right )^2}
,
\end{equation}

We study the small-world model with the distribution of random
bonds defined in Eq. (\ref{PP}) and a fixed positive long-range
coupling constant as in Eq. (\ref{dPF}). We  start in subsection
\ref{ferro} to study the ferromagnetic case with $J_0>0$ and
the location of the critical points for different values of $c$
and $d_0=1,2$ and 3. In subsection \ref{cf}  we analyze the
spin-spin correlation function above the critical temperature for
$d_0=2$. In subsection \ref{nonzerofield} for the special case
$J_0=0$ we study the magnetization and susceptibility at non-zero
external field above the critical temperature. In subsection
\ref{2orderJ0negative} we consider a one dimensional system with
$J_0$ negative where two second-order phase transitions are
predicted. In subsection \ref{1orderSGJ0negative} we analyze the
same one-dimensional model for couplings and connectivity such
that either two first-order phase transitions or a spin-glass
phase are predicted. Finally in subsection \ref{bimodalEA} we
study a two dimensional Edwards-Anderson model with added
long-range shortcuts.

\subsection{The Paramagnetic-Ferromagnetic line of critical points}
\label{ferro} From the susceptibility of the pure Ising
model in a hypercubic lattice of dimensionality $d_0$  and Eq.
(\ref{THEOg}) we obtain the location of the
P-F line of critical points in the c-T
plane. For $d_0=1$ we  use the known analytical expression for the
susceptibility, $\tilde{\chi}_0\left(\beta J_0,0\right)=\exp(2\beta J_0)$,
that applied to Eq. (\ref{THEOg}) reproduces the same formula of \cite{Niko} for the P-F and P-SG lines.
For higher dimensions we use
numerical results obtained from Metropolis Monte-Carlo
simulations.

\begin{figure}
\includegraphics{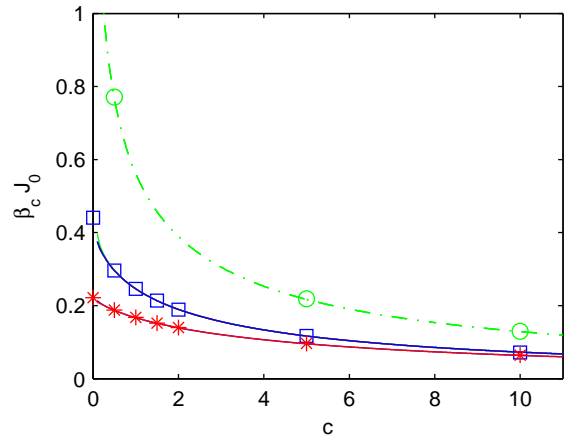}
\caption{\label{TcFormula}(Color online) Lines of critical points in the $\beta
- c$ plane for $d_0=1$ ($\bigcirc$) with $J/J_0=3/5$, and for
$d_0=2$ ($\square$) and $d_0=3$ ($\star$) with $J/J_0=1$ as
obtained from the pure model susceptibility and Eq.
(\ref{THEOg}). The symbols are estimates of the critical points
obtained from simulation and by using the Binder cumulant
intersection technique.}
\end{figure}

For $d_0=2$ the pure model susceptibility was determined
for several systems sides up to $L=128$ to check for finite-size
effects.  The pure model was simulated for 40
temperatures, in the range $0.1 <\beta J_0 <= \beta_{c0}
J_0=0.44068...$. For $d_0=3$ we studied systems of side $L=8$ and
$L=16$ also for 40 temperatures in the range $0.05 <\beta J_0 <=
\beta_{c0} J_0=0.2216546(10)$ \cite{Ising3d}. In all of the
simulations reported we neglected the first $10^5$ MCS/N and made
measurements in the remaining $10^6$ MCS/N steps. In Fig.
\ref{TcFormula} we plot the critical lines for $d_0=1$ with
$J/J_0=3/5$, and for $d_0=2$ and 3 with $J/J_0=1$. Note that in the
figure, for $d_0=2$ and $ 3$, there are several lines
corresponding to the use of susceptibility estimates obtained from
systems of different size. Nevertheless, in the scale of the plot
no finite size effects can be seen. This is a consequence of the
fact that only very near $c=0$ the solution of Eq. (\ref{THEOg})
uses values of the pure model susceptibility near the
critical point. Far from the critical point of the pure
model the susceptibility and consequently the estimates of the
critical line in the disordered model do not show finite-size
effects.

 In order to compare the predictions based on Eq. (\ref{THEOg}) we measured by direct simulation
 the location of the critical points for several values of the
 average connectivity, $c=0, 0.5, 1, 1.5, 2, 5$ and $10$ and the results, for each spatial dimension, are plotted in Fig.
 \ref{TcFormula}.
 The average over disorder was done by considering averages over 10 samples.

\begin{table}
\caption{(Color online) Results of Binder cumulant crossings for $d_0=2$ and
$J_0>0$. In the
 column $\beta_{L,L'}J_0$ we list the crossing inverse temperatures between the cumulants for systems of side $L$ with $L'=128$.
 In the column $\beta_c J_0$ we list the critical parameters obtained from Eq. (\ref{THEOg}). The crossing values of the cumulant are listed
 in the column $U^*_{L,L'}$. \label{tab:1}}
\begin{ruledtabular}
\begin{tabular}{llllll}
$c$ & $L$  & $\beta_{L,L'}J_0$ & $\beta_c J_0$ & $U^*_{L,128}$
\\ \hline\hline
 & 16 & 0.4410 & & 0.612 \\
0  & 32  & 0.4409 & $0.440686...$               &0.612\\
  & 64  & 0.4411 &               &0.615 \\ \hline
  & 16  & 0.2968&              & 0.33\\
  0.5 & 32 & 0.2964&  0.2963      & 0.30\\
      & 64 & 0.2960&              & 0.27 \\ \hline
      & 16  & 0.2466&              & 0.32 \\
  1.0 & 32  & 0.2461&  0.2461      & 0.28 \\
      & 64  & 0.2460&              & 0.28 \\ \hline
      & 16  & 0.2139&              & 0.30 \\
  1.5 & 32  & 0.2137&  0.2134      & 0.29 \\
      & 64 & 0.2136&              & 0.28 \\ \hline
      & 16 & 0.1894&              & 0.30\\
2.0   & 32 & 0.1891& 0.1897       & 0.27\\
      & 64 & 0.1888 &             & 0.24 \\ \hline
      &  16 & 0.1173&             & 0.28\\
5.0   & 32 & 0.1174 & 0.1167      & 0.29\\
      & 64 & 0.1173 &             & 0.27\\ \hline
      & 16 & 0.0730 &             & 0.28     \\
10.0  & 32 & 0.0730 &0.0728       &0.28      \\
      & 64 & 0.0731 &             & 0.30    \\ \hline
  \end{tabular}
\end{ruledtabular}
\end{table}

For the case $d_0=2$ we present in Table \ref{tab:1} detailed
numerical results. For the cumulant crossing $\beta_{L,L'}J_0 $
(third column) listed for several values of $c$ we estimate a
statistical error $0.0005$ which allows us to claim a good
agreement between the simulations and the theoretical prediction
obtained from Eq. (\ref{THEOg}) (fourth column). The values of the
cumulant at the critical point for the pure model ($c=0$) are
close to the value $0.61069$ calculated in \cite{CumulantIsing}.
The long-range links introduced by the disorder change the
universality class from 2d Ising to mean-field.

The mean-field value of the Binder cumulant at criticality for an
infinite system is predicted to be $
0.2705$\cite{Luijten1995,Parisi2,Luijten1999} which is close to
our estimates of Binder cumulant intersections listed in table
\ref{tab:1} (fifth column) for which we estimate an error equal to
$0.02$. In three dimensions we simulated only systems of side
$L=8$ and 16. For $c=0$ the two cumulants intersect at the value
$U_L=0.486$ close to the estimation $0.46521$ reported in
\cite{Ising3d}. For the other values of $c$ studied, $c=0.5, 1,
1.5, 2, 5$ and $10$ the intersection was measured near $U_L=0.31$.
In one dimension we studied only $c=0.5, 5$ and $c=10$ and also
the Binder cumulant intersections were found to be near $0.3$ for
intersections of $L=128,256,512, 1024$ with $L'=2048$.

\begin{figure}
\includegraphics{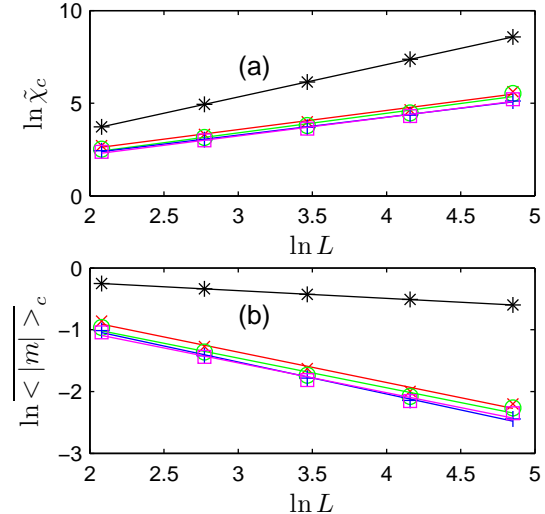}
\caption{\label{crit_exp}(Color online) In (a) we plot the susceptibility,
$\tilde \chi_c$, and in (b) the magnetization, $\overline{\left <
|m| \right>}_c$, at the critical point for the $d_0=2$ model with
$J/J_0=1$ as a function of the system side $L$. Both in (a) and
(b) the data are for $c=0 (\star), 0.5 (\times), 1 (\bigcirc), 1.5
(+), 2 (\square)$. }
\end{figure}

Furthermore, for $d_0=2$, we measured the scaling with system size
of the average value of the absolute value of the magnetization
$\overline{\left < |m| \right >}(\beta_c J_0)$ and the
susceptibility, $\tilde \chi(\beta_c J_0)$ at the critical point.
For a mean-field universality class these quantities are expected
to scale at criticality like $\overline{\left < |m| \right
>_c}\sim N^{-1/4}\sim L^{-d_0/4}$ and $\tilde \chi_c\sim
N^{1/2}\sim L^{d_0/2}$ where $N$ is the total number of spins
\cite{Luijten1995,Parisi2,Luijten1999}. In Fig. \ref{crit_exp} we
show these quantities plotted in bi-logarithmic scale as a
function of system size for the different $c$ values studied. For
the magnetization, the slope of the straight line fit for $c=0$ is
$-0.125(1)$  and for the susceptibility is $1.752(1)$, consistent
with the known exact exponents of the pure $d_0$=2 Ising model.
For $c=0.5, 1, 1.5$ and $2$ we got, respectively, for the
magnetization exponents $-0.49(3),-0.48(3),-0.52(2)$ and
$-0.48(3)$ close to the expected value $-0.5$; whereas for the
susceptibility we obtain the exponents $1.03(5), 1.06(7), 0.97(2)$
and $1.01(5)$, also close to the expected result for mean-field
behavior.

\subsection{Correlation functions above the critical temperature}
\label{cf} 
\begin{figure}
\includegraphics{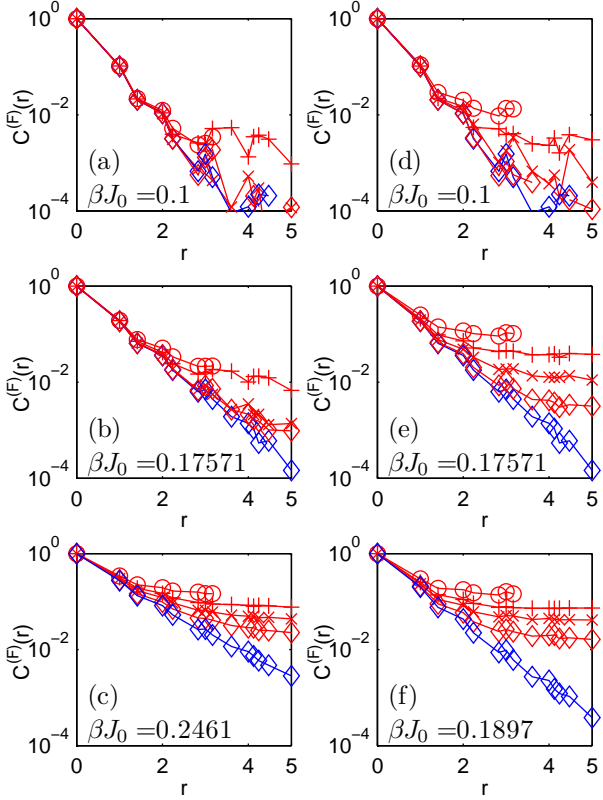}
\caption{\label{CF}(Color online) Correlation functions $C^{(F)}(r)$ as a
function of Euclidean distance $r$ for  $J=J_0$ and $c=1$ (in (a),
(b) and (c)) and $c=2$ (in (d), (e) and (f)) at different
temperatures, $\beta J_0=0.1$ ((a) and (d)), $0.17571$ ( (b) and
(e)). In (c) with $\beta_c J_0=0.2461$ and in (f) with $\beta_c
J_0=0.1897$  (the critical temperatures for $c=1$ and $c=2$,
respectively). The lower line (which is blue in the online version
of this paper) is $C^{(F)}(r)$ for the pure model $c=0$ and for a
system side $L=64$ ($\Diamond$). Data for four system sides are
plotted, $L=8$ ($\bigcirc$), $L=16$ ($+$), $L=32$ ($\times$),
$L=64$ ($\Diamond$).}
\end{figure}

From Eq. (\ref{THEOh}) we see that the effective field theory, in the P region
and zero external field, predicts the spin-spin correlation
function of the random model $C^{(F)}$ to be, in the thermodynamic limit, equal to the
correlation function of the pure model calculated at the same
temperature. In order to check this result we calculated, from
simulation,
\begin{eqnarray}
\label{CFf} C^{(F)}(r)=\overline{\left < \sigma_0 \sigma_r \right
> },
\end{eqnarray}
where $\sigma_0$ is an arbitrary spin and  $\sigma_r$ is one spin
at Euclidean distance $r$ from the spin $\sigma_0$, measured on
the lattice $\mathcal{L}_0$. We considered the case $d_0=2$
with $J=J_0$ and $c=1$ and $2$ and the inverse temperatures
$\beta J_0=0.1, 0.17571$. These temperatures are above the
critical temperatures for the two values of $c$ studied. We
studied also the correlation function at the critical inverse
temperatures $\beta_c J_0=0.2461$ and  $\beta_c J_0=0.1897$ for $c=1$
and $c=2$, respectively (see Table \ref{tab:1}). These
calculations were done for system sides $L=8,16, 32$ and $64$. In
Fig. \ref{CF} we can see that, as the system size increases, the
curves $C^F(r)$, for $c>0$, approach the data points for $c=0$
calculated for a system of size $L=64$. Note that, as we approach
the critical temperature, the correlation function at large $r$
reaches a finite value that decreases as the system size
increases. This finite constant is just due to the finite size of
the system and it is predicted by the theory to vanish as $1/N$,
in the thermodynamic limit, but at the same time it is responsible
for the divergence, with system size, of the susceptibility at the
critical point (see Eq. (33) and subsequent comments of Ref. \cite{SW}).

\subsection{Susceptibility  and Magnetization at non-zero external field above the critical temperature}
\label{nonzerofield}
In the P phase at zero external
magnetic field the effective field theory prediction for the
susceptibility is exact, consistently with the exact predictions
for the critical temperatures. The question remains whether the
susceptibility prediction is a good approximation for non-zero
field. To verify this we made simulations for the case $J_0=0$
with a positive long-range coupling  (as in Eq. (\ref{dPF})) (in
other words the simplest version of Viana-Bray model). For this
particular case the magnetization is predicted to be given by the
solution of the following equation
\begin{eqnarray}
\label{VianaBrayM} m=\tanh\left( c~ m~ \tanh(\beta J) + \beta
h\right )
\end{eqnarray}
and the susceptibility (divided by $\beta$) is given by,
\begin{eqnarray}
\label{VianaBrayChi} \tilde \chi(\beta J, \beta h)=\frac{1-m^2}{
1-c (1-m^2) \tanh \beta J}
\end{eqnarray}

\begin{figure}
\includegraphics{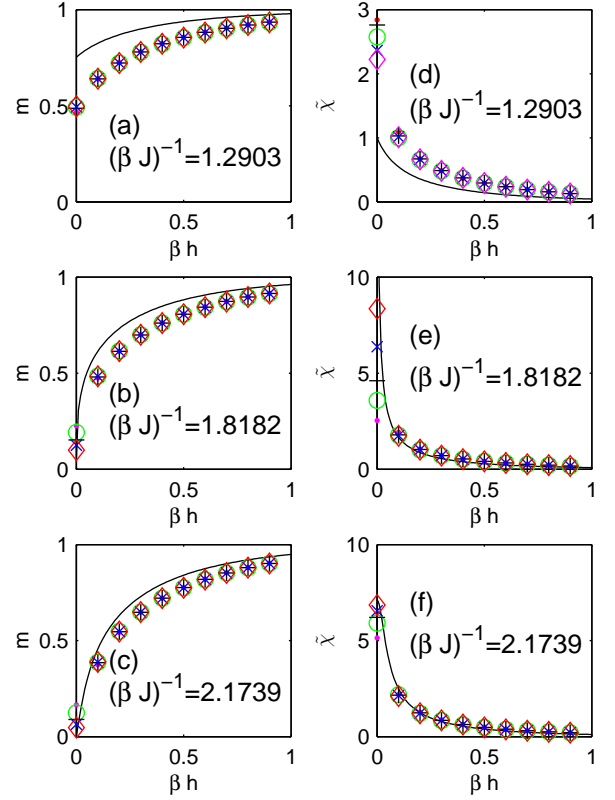}
\caption{\label{VianaBray}(Color online) Magnetization ( (a), (b) and (c) ) and
Susceptibility ( (d), (e), (f)) for the Viana-Bray model and $c=2$
for three temperatures as a function of the external magnetic
field. The lines are the theoretical predictions ( see Eqs.
(\ref{VianaBrayM}) and (\ref{VianaBrayChi})) and the data points
are simulation results for $N=128 (.), 256 (\circ), 512 (+), 1024
(\times)$ and $2048 (\Diamond)$. For the case studied here the
critical temperature is, $(\beta_c J)^{-1}=1.8205$. }
\end{figure}

In Fig. \ref{VianaBray} we compare the results of the above
predictions for the magnetization  and susceptibility with
simulation results for $c=2$. For averaging purposes we considered
$10$ samples. The plots correspond to three temperatures $(\beta
J)^{-1}=1.2903$ (first row), $(\beta J)^{-1}=1.8182$ (middle) and
$(\beta J)^{-1}=2.1739$ (bottom). The critical temperature is
$(\beta_c J)^{-1}=1.8205$. By construction, in the limit of strong
field the magnetization prediction becomes exact and similarly in
the limit of small field above the critical temperature. In the
intermediate field range we see that the magnetization prediction
and simulation results in general do not agree. However, above the
critical temperature, the susceptibility obtained from simulation
and the theoretical prediction given by Eq. (\ref{VianaBrayChi})
are very close to each other over the full range of field values
studied. Note that for $(\beta J)^{-1}=1.8182$, close to the
critical temperature, the simulation susceptibility shows, as
expected, a strong finite-size effect at zero field.

It is worth to observe that with respect to our effective field
theory, the Viana-Bray model represents the worst, \textit{i.e.},
the most difficult, case. The theory in fact, by construction,
takes exactly into account all the effects due to the short-range
couplings and to the short-loops present in the given lattice
$\mathcal{L}_0$, and the greater is $d_0$, the greater is the
level of accuracy of the theory also out of the P
region (at least in the absence of frustration), while in the
Viana-Bray model topologically we have $d_0=0$.

\subsection{Negative short-range coupling and second order phase transitions}
\label{2orderJ0negative} 
\begin{figure}
\includegraphics{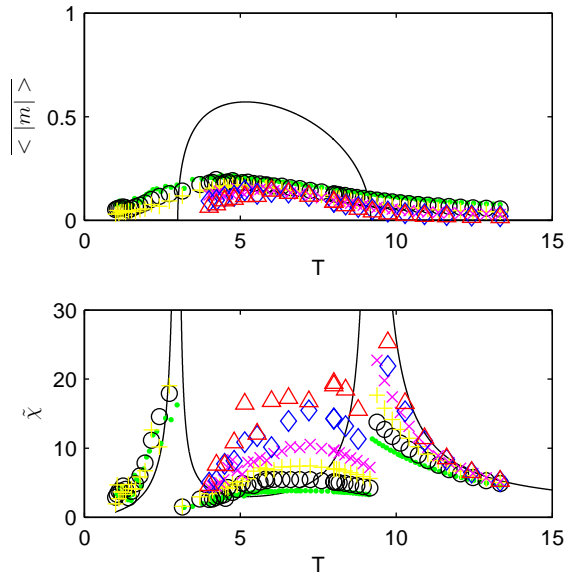}
\caption{\label{MChiJ0neg}(Color online) Magnetization (top) and susceptibility
(bottom) as a function of temperature for a $d_0=1$ system with
$J_0<0$. The average connectivity is $c=1.4$ and $J_0=-0.5$,
$J/J_0=-20$. The line is the theoretical prediction and the data
points correspond to different system sizes, $L=512 (.), 1024
(\circ), 2048 (+), 4096 (\times), 8192 (\diamond), 16384
(\triangle)$.}
\end{figure}

In the case $J_0<0$ the theory allows for
the occurrence of two second-order phase transitions. At low and
high temperatures the system is disordered and in the intermediate
temperatures a ferromagnetic phase arises. The simulations confirm
this phase diagram picture. We made simulations at $d_0=1$, for
$J_0=-0.5$, $J/J_0=-20$ and $c=1.4$. We made averages over
disorder by considering 50 samples and we studied system sizes
$L=512, 1024, 2048, 4096, 8192$ and $16384$. In Fig,
\ref{MChiJ0neg} we plot the magnetization and the susceptibility
as a function of temperature, $T$. The two critical points are
predicted to occur at $T_{c,1}=2.985$ and $T_{c,2}=9.207$. The
predicted values of the magnetization in the intermediate
temperature range, $T_{c,1}<T < T_{c,2}$, are  different from the
simulation results (see Fig. \ref{MChiJ0neg}). However, the
theoretical predicted susceptibility, in both P phases,
is very close to the simulation susceptibility approaching each
other as the system size increases.

\begin{figure}
\includegraphics{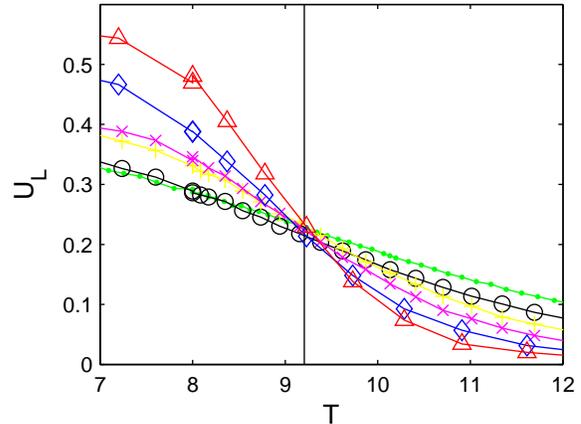}
\caption{\label{BinderCriticalJ0negative}(Color online) Binder cumulant $U_L$ as
a function of temperature for a $d_0=1$ system with $J_0<0$. The
average connectivity is $c=1.4$ and $J_0=-0.5$, $J/J_0=-20$. The
vertical line is the theoretical prediction for the critical
temperature and the data points correspond to different system
sizes, $L=512 (.), 1024 (\circ), 2048 (+), 4096 (\times), 8192
(\diamond), 16384 (\triangle)$.}
\end{figure}

Note that this model is a frustrated system so that large fluctuations and strong finite size
effects are present, especially close to the lower temperature
critical point where we do not perform high precision simulation
as it requires averaging over a large number of samples. We have
studied, in detail, the high temperature critical point where we
applied the cumulant crossing technique.
The intersection temperatures of the cumulants for $L=512, 1024,
2048, 4096$ with the system of size $L=16384$ were, $9.24, 9.35,
9.23, 9.32$, respectively, from which we can estimate a critical
temperature equal to $9.29(6)$. This estimate is close but
slightly higher than the predicted critical temperature,
$T_{c,2}=9.207$. However, considering the statistical error and
the finite size corrections we cannot exclude a convergence toward
the theoretical value in the thermodynamic limit. The
corresponding intersection values of the cumulant were, $0.23,
0.21, 0.23, 0.21$. These values are smaller than the values in the
range $ 0.27-0.3$ that we measured in section \ref{ferro} for
$J_0>0$. In the Fig. \ref{BinderCriticalJ0negative} we plot the
Binder cumulant as a function of temperature for the system sizes
studied.

\subsection{Negative short-range coupling, first-order  and spin-glass phase transitions}
\label{1orderSGJ0negative} We considered the case for $c=10$,
$J_0=-0.9$ and $J=0.5$ at $d_0=1$.
From the theory it turns out that for temperatures above $2.38$
only the zero magnetization solution is stable but for
temperatures lower than this value there are always two stable
solutions, one with non-zero magnetization and another with zero
magnetization.  For temperatures $T<2.34$ the nonzero
magnetization solution has the lower free-energy so that a
first-order phase transition is predicted at this temperature. The
theory also predicts a possible spin-glass phase transition at a
temperature, $T_{c,SG}=1.88$. Note that the theory always predicts
continuous spin-glass phase transitions.

\begin{figure}
\includegraphics{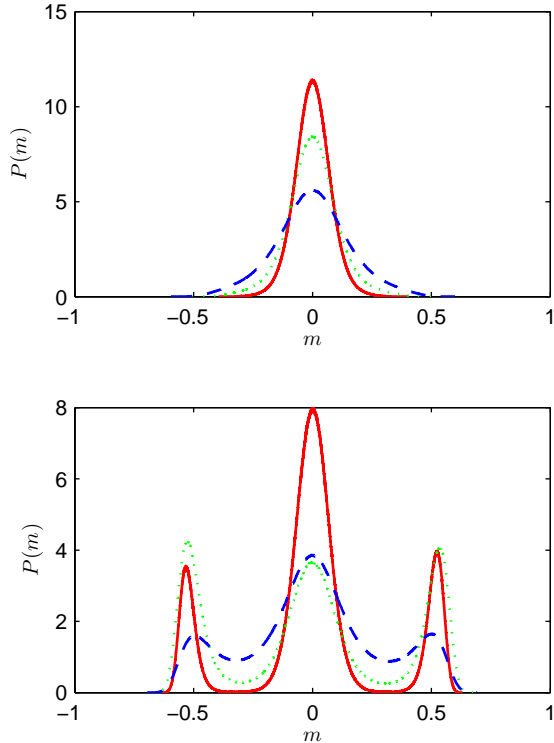}
\caption{(Color online) Simulation results for the  magnetization probability
distribution, $P(m)$, for the model with $c=10$, $J_0=-0.9$
and $J=0.5$ at $d_0=1$. In (a) we plot curves for $L=8192$ (solid
line), $L=4096$ (dotted line) at $T=2.05$ and $L=2048$ (dashed
line) at $T=2$ and in (b) we plot the corresponding data for
$L=8192$ and $L=4096$ at $T=1.86$ and $L=2048$ at $T=1.83$.
\label{pmFirstOrder} }
\end{figure}

We made simulations for systems of size $L=512$, $1024$,
$2048$, $4096$ and $8192$ and as before we neglected $10^5$  MCS/N
for equilibration purposes and we made measurements for $10^6$
MCS/N. The averaging over disorder was made by considering 50
samples. The results show a first order phase transition occurring
at slightly lower temperatures than the one predicted by the
theory. The probability distribution of the magnetization clearly
exhibits (see Fig. \ref{pmFirstOrder}) the behavior characteristic
of first-order phase transitions \cite{vollmayr1993} namely the
emergence, at temperatures close to the transition, of two maximum located
at symmetric nonzero values together with a third
maximum near zero magnetization.

\begin{figure}
\includegraphics{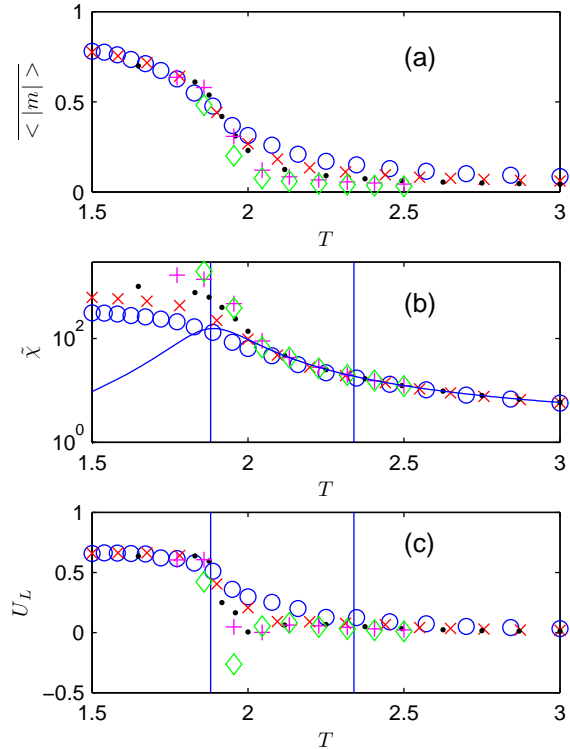}
\caption{(Color online) Simulation results for the average magnetization (a),
magnetic susceptibility (b) and Binder cumulant $U_L$ (c) as a
function of temperature for the model with $c=10$,  $J_0=-0.9$ and
$J=0.5$ at $d_0=1$. In (b) we have used  $\tilde \chi=N
\overline{<m^2>}$ at any temperature. The vertical lines are at
$T_c=2.34$ and $T_{c,SG}=1.88$. In (b) the line is the theoretical
susceptibility for the zero magnetization solution. The simulation
data points are for $L=512 (\bigcirc), 1024 (\times), 2048
(.),4096 (+), 8192 (\diamond)$. \label{FirstOrderTrans} }
\end{figure}

In Fig. \ref{FirstOrderTrans}(a) we show the simulation results
for the average magnetization, the magnetic susceptibility (b),
and the magnetization Binder cumulant (c). The simulation
susceptibility follows very closely the  zero magnetization
theoretical susceptibility up to temperatures lower than the
predicted $T_c$. Note that, in the case under study, the zero
magnetization solution is a stable solution at any temperature and
the theoretical prediction of the phase transition is based on
comparison of the value of free-energies of the solutions. The
theory does not predict correctly the location of the transition
since values of the free-energy of the non-zero magnetization
solution are not given exactly by the theory. The simulation
Binder cumulant shows, near the transition temperature, the
expected increasingly negative values, as the system size
increases \cite{vollmayr1993}. Interestingly, the simulation data
give a transition temperature close to the theoretically predicted
spin-glass transition temperature.

\begin{figure}
\includegraphics{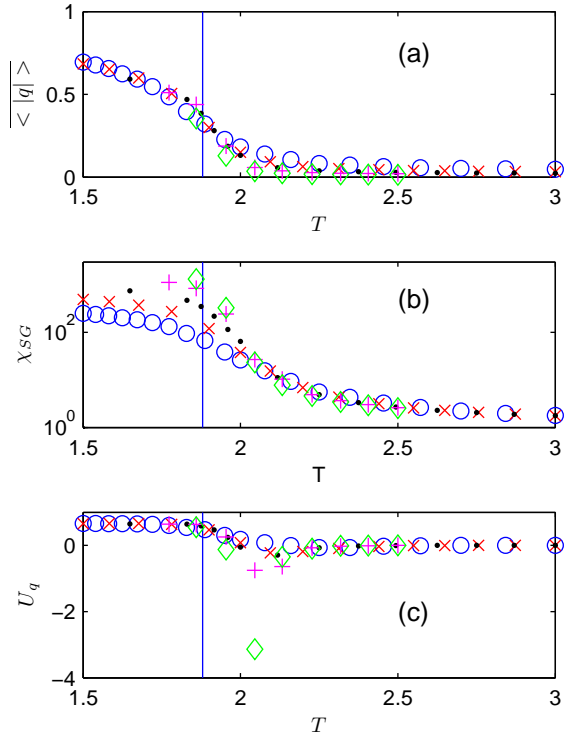}
\caption{(Color online) Simulation results for the average overlap order
parameter, $\overline{<|q|>}$ (a), $\tilde \chi_{SG}=N
\overline{<q^2>}$ (b) and Binder cumulant, $U_q$ (c), as a
function of temperature for the model with $c=10$, $J_0=-0.9$ and
$J=0.5$ at $d_0=1$. The data points are simulations for $L=512
(\bigcirc), 1024 (\times), 2048 (.),4096 (+), 8192 (\diamond)$.
\label{FirstOrderTransQ} }
\end{figure}

We also studied the overlap order parameter distribution and the
results are shown in Fig. \ref{FirstOrderTransQ}. Here, also, the
behavior expected for  a first-order transition temperature was
observed. The negative value minima of the Binder cumulant for the
overlap order parameter are steeper and occur at slightly higher
temperatures, for the same system sizes, as compared with the
corresponding quantity for the magnetization.

Quite interestingly for $c=4.5$ and the  values of the coupling
constants, $J_0=-0.9$ and $J=1$, the theory predicts the absence
of ferromagnetic phase transitions and only a continuous P-SG
phase transition located at $T_c^{(SG)}=2.29$.

\begin{figure}
\includegraphics{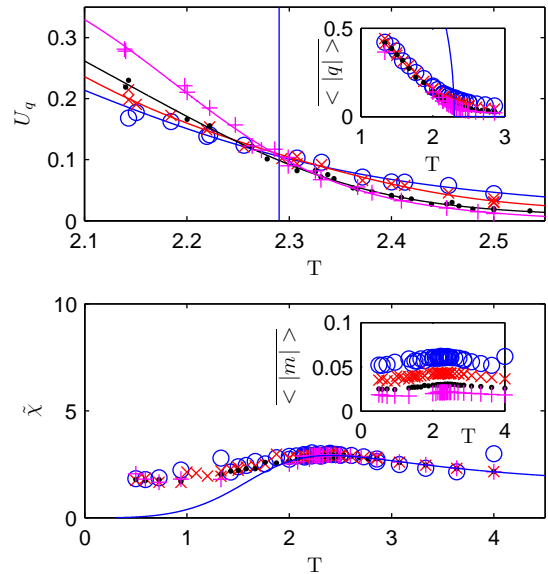}
\caption{(Color online) Results for the model with $c=4.5$,  $J_0=-0.9$ and
$J=1$ at $d_0=1$. In the top plot  we show the temperature
dependence of the Binder cumulant for the overlap order-parameter.
The lines are fits made for each system size. The vertical line is
the location of the predicted spin-glass critical temperature. The
inset shows the average values of the overlap order parameter
together with the effective field theory prediction for this
quantity.  In the bottom plot we show the temperature dependence
of the magnetic susceptibility. The inset shows the average
magnetization as a function of temperature. For all the plots the
data points correspond to simulations for systems of sides, $L=512
(\bigcirc), 1024 (\times), 2048 (.),4096 (+)$.
\label{J0negativeSG}}
\end{figure}

The obtained simulation results  are shown in Fig.
\ref{J0negativeSG}. The number of samples was 100 and the
simulation times considered here were the same as for the case
$c=10$. In the top plot we see that $U_{q,L}$ for different system
sizes intersects very near the theoretically predicted critical
temperature. The inset of the top plot shows the average overlap
order parameter that decreases with  system size above the
spin-glass critical temperature. In the lower plot we see that the
magnetic susceptibility is almost independent of system size and
it agrees with the theoretical prediction for temperatures above
the spin-glass phase transition while it deviates at lower
temperatures.

\subsection{Bimodal Edwards-Anderson model}
\label{bimodalEA}

The bimodal Edwards-Anderson  model\cite{edwards} is a disordered
spin model where the nearest neighbor coupling $J_0$ of Ising type
spins has the bimodal distribution in (\ref{J0bimodal}). Here, we
consider the two dimensional square lattice version of the model
with additional long-range couplings. The pure
model (without long range shortcuts) is known to show a P-SG
phase transition only at zero temperature being the lower critical
dimension of the model equal to two \cite{Hartmann,katzgraber}.

\begin{figure}
\includegraphics{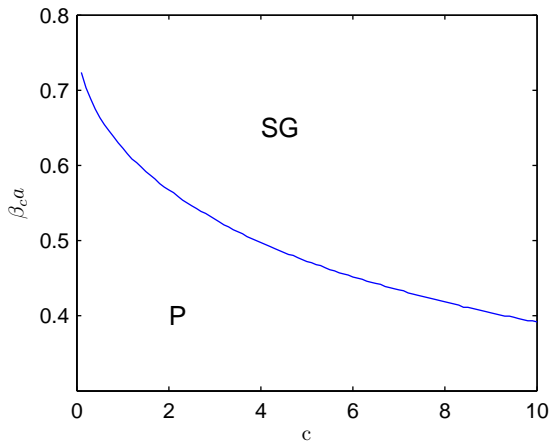}
\caption{\label{EAmodelPhasediag}(Color online) Phase diagram for the $d_0=2$
Edwards-Anderson model with long-range connections and coupling
constant $J=1$ . The curve is the value of $\beta_c a$ for the
SG phase transition as obtained from equation (\ref
{THEOgen}) and our numerical results for the pure Ising
model susceptibility for a system of side $L=128$. }
\end{figure}

To apply the effective field theory, for each phase $\Sigma=$F or
SG, we have to consider the pure Ising model magnetization
(\ref{THEOa}) and susceptibility (\ref{THEOchie}) calculated by
using the definitions of the effective long- and short-range
couplings $J^{(\Sigma)}$ and $J_0^{(\Sigma)}$, from Eqs.
(\ref{THEOb})-(\ref{THEOdb}), respectively.
By using the numerical two-dimensional Ising model magnetic
susceptibility, we can obtain the expected location of the
spin-glass phase transition. The result of this calculation is
shown in Fig. \ref{EAmodelPhasediag}. We stress that theory
predicts that the inclusion of an arbitrary small number of
long-range shortcuts in the Edwards-Anderson model leads always to
a finite temperature phase transition. In fact, in the limit of an
infinitesimal addition of short-cuts, the theory predicts a P-SG
transition at the finite value given by $\beta_c
\,a=\tanh^{-1}[\sqrt{\tanh(0.44068...)}]=0.7642...$,where
$0.44068...$ is the critical inverse temperature of the regular
two-dimensional Ising model with a unitary positive coupling. This
implies that the EA model with short-cuts, in the limit $c\to
0^{+}$, is not equivalent to the original EA model without
short-cuts. In other words, the EA model is not thermodynamically
stable under graph noise.

\begin{figure}
\includegraphics{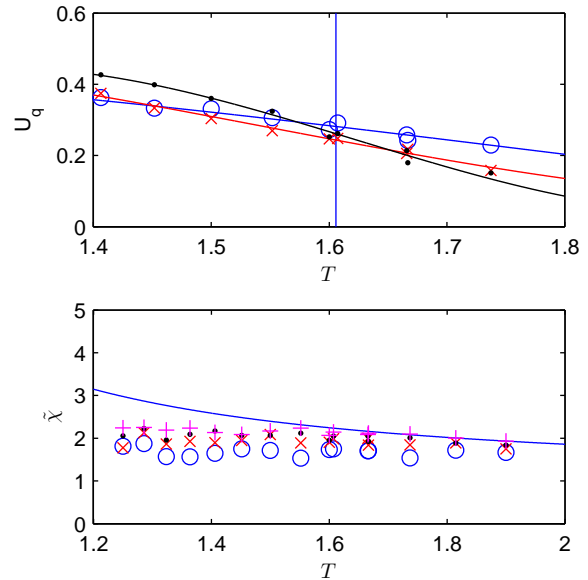}
\caption{\label{EAmodel}(Color online) Results for the $d_0=2$ Edwards-Anderson
model with long-range connections with $a=J=1$ at $c=1$. The top
plot shows values of the Binder cumulant as a function of
temperature and the bottom plot shows the temperature dependence
of the magnetic susceptibility. The simulations were done for
$L=8$ ($\bigcirc$), $L=16$ ($\times$), $L=32$ (.) and $L=64$
($+$). In the top plot the vertical line is the theoretical
prediction of the spin-glass critical temperature and the lines
are polynomial fits of the Binder cumulants for each system size.
In the lower plot the line is the theoretical prediction for the
magnetic susceptibility. }
\end{figure}

We made simulations for the EA model with $a=1$, $J=1$ and $c=1$
to compare with the predictions of the theory. The average over
disorder was done by considering 100 samples. We neglected the
first $10^5$ MCS/N for equilibration and made measurements for
$10^6$ MCS/N. For $c=1$ the expected critical temperature is
$T_{c}^{(SG)}=1.6057$. In Fig. \ref{EAmodel} (top panel) we
obtained crossings of the overlap parameter Binder cumulant for
system sizes $L=8, 16$ and $32$. The statistical error of the
Binder cumulant, as expected, increases with system size and we
excluded from the Binder cumulant intersection calculations the
data for $L=64$. Our numerical estimate of the critical
temperature is $T_{c}^{(SG)}=1.6(1)$ still consistent with the
expected critical temperature.

Since, for an Ising model with zero coupling and zero external
magnetic field $\tilde \chi_0=1$, the prediction for the
magnetization and the magnetic susceptibility is $<|m|>=0$ and
$\tilde \chi= [1-c\tanh(\beta J)]^{-1}$. In Fig. \ref{EAmodel}
(lower panel) we plot the magnetic susceptibility simulation
results together with this theoretical prediction. We see that in
the P phase the numerical estimates of the susceptibility approach
the theoretical curve as the system size increases.

\section{Conclusions}
In this work we have compared the predictions of an effective
field theory for several Ising models on small world networks with
Monte Carlo simulation results. All the predictions of the theory,
where it is known to be exact, namely in the P region
at zero external field, were confirmed by the simulation results.
In particular we have checked the  critical surfaces $(T,c)$, the
temperature dependence of the susceptibility, and the spin-spin
correlation function.

Furthermore, for the simplest version of the model, \textit{i.e.},
the Viana-Bray model (where $d_0=0$), we studied the effect of a
non-zero external field. Although the theory is not exact in this
case, there is a reasonable agreement between the predicted field
dependence of the magnetization and susceptibility specially above
the critical temperature. For the case of $J/J_0=-20$ and $c=1.4$
at $d_0=1$ we verified the existence of two second order phase
transitions. A good agreement between the theoretical temperature
dependence of the susceptibility in the P phases and
simulation results was observed. The location of the high
temperature critical point was explicitly verified using the
cumulant crossing technique.

For a one-dimensional model with $J_0=-0.9$, $J=0.5$ and $c=10$,
we observed a first order phase transition as predicted by the
theory but at a lower temperature. This may be a consequence of
the fact that in this case, unlike the second-order phase
transitions cases, the zero magnetization solution does not become
unstable at the transition temperature and - consistently with the
fact that the theory does not give exact results out of the pure P
regions - the critical point obtained as the point where the two
free-energy values equal is not exactly predicted. Quite
interestingly, we find that also the P-SG phase transition turns
out to be first-order and, furthermore, its critical point seems
to coincide with the theoretical one.

With the couplings $J_0=-0.9$, $J=1$ and a  smaller average
connectivity, $c=4.5$ the theory predicts that only a P-SG transition is present.
We studied by simulation this
spin-glass critical behavior and we found the critical temperature
very close to the theoretical prediction.

We also introduced a model not previously studied, the two
dimensional Edwards-Anderson model with added long-range
shortcuts, where we confirmed that even an infinitesimal inclusion
of shortcuts makes the spin-glass phase transition to occur at a
finite non-zero temperature. In other words, as the theory
predicts, we find that the two dimensional Edwards-Anderson model
is not thermodynamically stable under graph-noise.

The class of disordered models for which the theory is applicable
is very wide and its application just relies on the  availability
of numerical or analytical results  for the susceptibility of non
disordered models in arbitrary $d_0$ dimensions. When there is
strong frustration, simulations are difficult to perform requiring
large simulation times and averages over many samples. Our results
clearly confirm the usefulness of the effective field theory
proposed in \cite{SW} by giving accurate predictions for the
models phase diagram. The possibility to improve the theory out of
the P region opens a new interesting challenge.


\begin{acknowledgments}
This work was supported by the projects SOCIALNETS and  FCT (Portugal) PTDC/FIS/71551/2006.
We thank M. Barroso for the administration of the computational facilities
where the simulations were done.
\end{acknowledgments}

\newpage



\end{document}